\documentclass[%
 reprint,
 amsmath,amssymb,
 aps,
]{revtex4-2}

\usepackage{graphicx}% Include figure files
\usepackage{dcolumn}% Align table columns on decimal point
\usepackage{bm}% bold math
\usepackage{ulem}
\usepackage{xcolor}

\begin{document}

\preprint{APS/123-QED}

\title{Granular drag and lift force on a flexible fiber}

\author{Antonio Pol}
 %\altaffiliation[Also at ]{Physics Department, XYZ University.}
 \email{antonio.pol@inrae.fr}

\affiliation{IATE, Univ. Montpellier, INRAE, Institut Agro, F-34060 Montpellier, France}%

\author{Sara Storti}
\affiliation{IATE, Univ. Montpellier, INRAE, Institut Agro, F-34060 Montpellier, France}%
\affiliation{Department ICEA, University of Padova, Via Ognissanti 39, 35129 Padova, Italy}

\author{Fabio Gabrieli}
\affiliation{Department ICEA, University of Padova, Via Ognissanti 39, 35129 Padova, Italy}

 %\altaffiliation[Also at ]{Physics Department, XYZ University.}
 %\email{antonio.pol@inrae.fr}
%\affiliation{UMR IATE?, University of Padova? ..,}%}%

\date{\today}% It is always \today, today,

\begin{abstract}
In this work, we investigate the forces acting on a flexible fiber dragged through a granular bed. Using discrete element simulations, we observe that, after a sufficiently large displacement, the system reaches a steady state in which both the fiber’s shape and the forces acting on it become, on average, constant. Under these conditions, we identify two characteristic lengths that describe the fiber's shape and propose unique scaling laws for the drag and lift forces, valid across a wide range of fiber flexibilities, from highly deformable to nearly rigid, based on these lengths. We highlight that the fiber-grains interaction is governed by a single dimensionless elastogranular parameter, defined as the ratio of the fiber’s elastic properties to the granular pressure. Finally, we demonstrate that both the forces and the characteristic lengths can be expressed solely as functions of this dimensionless parameter. Our findings offer a fundamental insight into the behavior of a flexible fiber interacting with a granular medium.
\end{abstract}

\maketitle

%\tableofcontents

\section{Introduction}\label{sec:intro}

Understanding how deformable fibers interact within granular environments is essential in various contexts, such as plant biomechanics (e.g., the growth and anchoring of plant roots), geotechnical engineering (e.g., fiber-reinforced soils), and material manufacturing (e.g., enhancing mechanical properties). In these cases, the emergent behavior at the macroscale results from fiber–grain interaction, which are governed by the evolution of the fibers shape and orientation under the constraints imposed by the granular phase.

The existing literature on the subject comes mainly from the geomechanical community and prevalently focuses  on the mechanical response of fiber-reinforced materials under compression or shear \cite{gray1983,maher1990,ahmad2010,diambra2010,ibraim2012,shukla2017}. In addition, recent numerical works \cite{yang2021,li2022} have addressed the impact of fibers orientation, stiffness and spatial distribution on the mechanical response of fiber reinforced granular media.
These works highligthed the enhancement of the mechanical strength of granular soils with the addition of fibers, but a systematic investigation of the impact of the fibers properties (e.g. length and stiffness) on the rheology of the mixture was not conducted.
Only recently, Wierzchalek et al.~\cite{wierzchalek2025} provided a physical insight into the rheology of grain-fiber systems. They experimentally tested ideal mixtures of glass beads and fibers of controlled properties and showed that the effective friction coefficient of the mixture increases linearly with the fibers' fraction and exponentially with the fibers' length.  

Despite recent progress in understanding grain–fiber mixtures, the fundamental knowledge of the behavior at the scale of a single fiber remains limited. In an attempt to fill this gap, %we investigate the interaction between a single fiber and a granular medium. 
we focus on a simple yet illustrative case: a fiber being dragged through a static granular layer. This configuration allows us to explore the effects of the interaction between the fiber and the granular medium, i.e. the elastogranular interaction, specifically the forces acting on it and the evolution of its geometrical configuration under those forces.
Interesting examples highlighting the rich behavior arising from elastogranular interaction are: the buckling/bending mechanics of an immersed flexible rod \cite{mojdehi2016,seguin2018,schunter2018,algarra2018} and the emergence of self-standing elastogranular structures \cite{fauconneau2016,guerra2022}. 

In the picture given above, the fiber can be seen as a peculiar, deformable intruder that moves inside a granular bed. The forces on an intruder have received great attention from the granular community \cite{albert1999,albert2001,albert2001stick,takehara2010,costantino2011,ding2011drag,potiguar2013,hilton2013,takehara2014,faug2015,seguin2016,artoni2019drag,seguin2022,carvalho2024drag}. In fact, these forces have significant implications for various practical problems, including mixing and segregation, impact and penetration, and animal or robotic locomotion in granular environments.
For rigid objects moving at constant velocity $v$ in a granular bed two regimes have been identified for the drag force depending on a Froude number $Fr=v/\sqrt{g h}$, where $g$ is the gravity and $h$ the burial depth of the object \cite{faug2015}. %(force acting opposite to the relative motion of the object with respect to the surrounding granular medium).
For $Fr\ll1$ the drag force is independent of the velocity $v$ and scales with the hydrostatic pressure $\rho g h$, with $\rho$ the density of the granular medium ($\rho=\phi \rho_g$ with $\phi$ the volume fraction of the medium and $\rho_g$ the grain density) \cite{albert1999,albert2001,costantino2011,hilton2013,seguin2022}. For $Fr\gg1$ instead, the drag force scales quadratically with the object velocity and follows a kinetic pressure scaling $\rho v^2$ \cite{takehara2010,potiguar2013,takehara2014,seguin2016}. In both cases, the drag force is given by a characteristic pressure, either hydrostatic or kinetic, multiplied by a characteristic section: the projection of the object along the direction of drag.
However, in our case, defining a characteristic geometrical quantity is not straightforward, as the shape of the fiber can evolve during the drag experiment to adapt to constraints imposed by the granular medium. Moreover, there are evidences that, even in the simple case of a rigid rod, the forces acting on the rod are strongly dependent on the orientation of the rod with respect to the drag direction \cite{ding2011drag,zhang2014}. 
Compared to the drag force on rigid intruders, the lift force remains less well understood. For a spherical body, Guillard et al. \cite{guillard2014} observed that the lift force is induced by a pressure gradient and follows a buoyancy-like scaling, with a lift coefficient that depends on burial depth and tends to saturate at large depths relative to the object size. Alternatively, Ding et al. \cite{ding2011drag} reported that, for drag experiments near the free surface, the lift force varies linearly with burial depth, similar to the drag force. They also observed that the lift force exhibits a remarkable non-monotonic dependence on object size and is influenced by the local orientation of the object's surface. Finally, Potiguar and Ding \cite{potiguar2013} showed that the lift force does not vary monotonically with drag velocity or intruder size, and that its behavior is strongly influenced by the shape of the intruder.

In this work, we use particle-based simulations aiming to answer the following questions:
(i) how does a flexible fiber evolve when dragged in a granular medium? and (ii) what are the forces that act on the fiber?

The paper is organized as follows. The numerical methodology is briefly described in Sec.~\ref{sec:methodology}. In Sec.~\ref{sec:num_results}, we present the results and propose scaling laws for the drag and lift forces. We devote Sec.~\ref{sec:discussion} to a discussion of these results, and we show that the behavior of the fiber is governed by a single dimensionless elastogranular number. Finally, in Sec.~\ref{sec:conclusion}, we summarize our findings and present future perspectives.

\section{Methodology}\label{sec:methodology}
\begin{figure}%[h!]
\includegraphics[width =.95 \columnwidth]{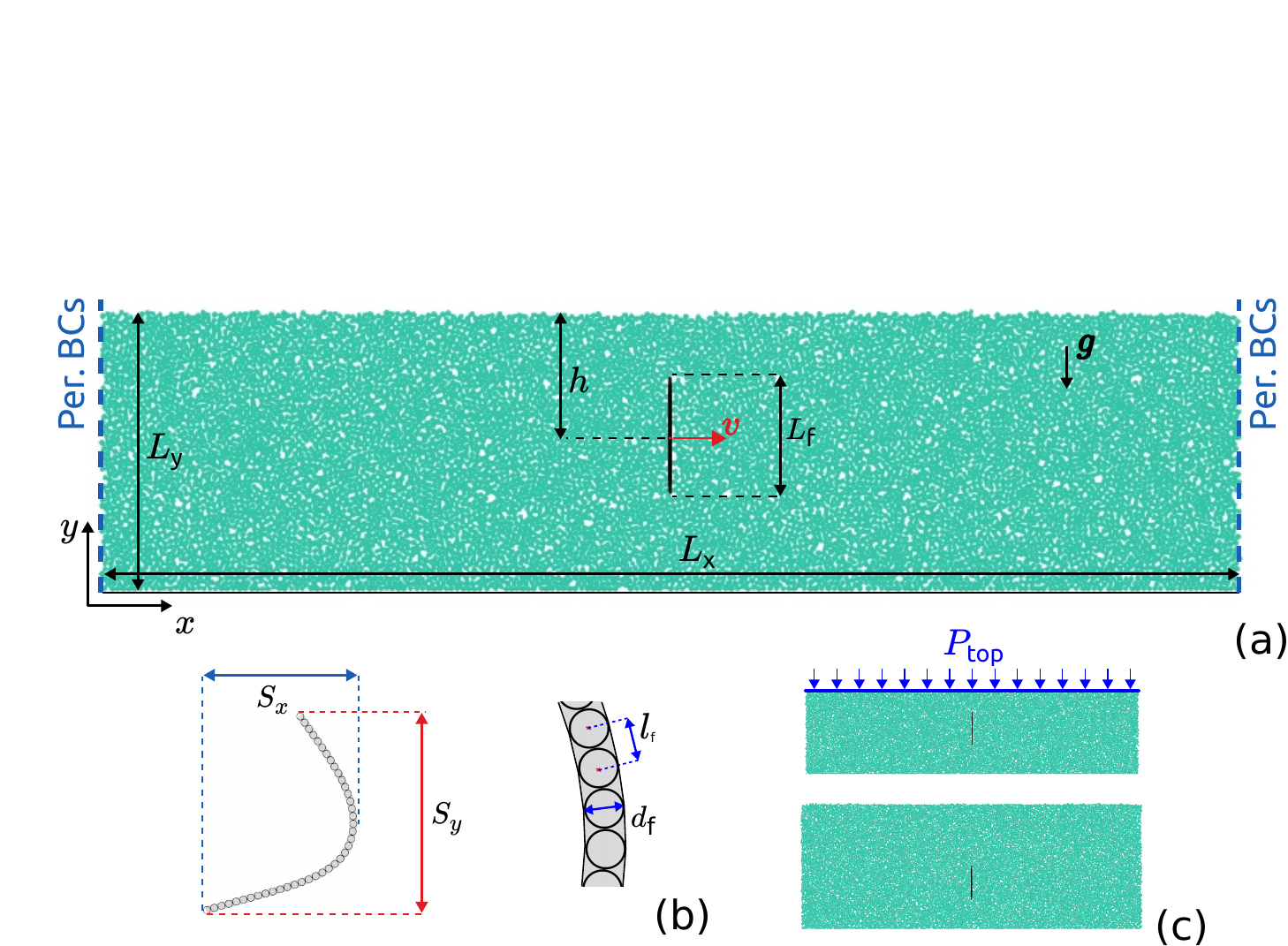}
\caption{(a) Typical geometry of the numerical setup. The granular bed  has dimensions $L_x\times L_y$. The fiber underformed length is $L_f$. The fiber is dragged inside the bed with a constant velocity $v$ along the $x$-direction. (b) Scheme of the deformed fiber with indication of the projections on the drag $S_x$ and orthogonal-to-drag $S_y$ directions, the diameter $d_f$ of the fiber and the distance between two nodal particles $l_f$. (c) Sketch of the configuration with a top pressure (top) and a thicker granular layer (bottom).}
\label{fig:fig0}
\end{figure}

We use discrete element simulations (DEM) to study the forces acting on a flexible fiber dragged in a granular bed. Simulations are performed with the open source code YADE \cite{Yade2021}. 
The granular bed is composed of slightly polydisperse spherical particles of diameter $d \pm 0.1d$ and particle density $\rho_g$ is set equal to $6/\pi$ in order to have unitary mass $m$. The system is subjected to gravity $g$ along the $y$-direction.
The granular bed has dimension $L_x \times L_y$ and only a single layer of particles is present along the out-of-plane direction (Fig.~\ref{fig:fig0}a). The out-of-plane translational and rotational degrees of freedom of the particles are blocked in order to study an equivalent of a 2D system. Periodic boundary conditions are applied along the $x$-direction, while in the $y$-direction, the granular bed is confined by a fixed planar wall at the bottom, with a free-to-deform top surface.
%---------------------------------------------
%---------------------------------------------

The fiber is modeled as a chain of cylindrical elements with rounded edges of diameter $d_f=d/2$ and length $l_f\approx d_f$ (a small offset of $10^{-3}d_f$ is used for numerical reasons), for a total length of the fiber $L_f$ (Fig.~\ref{fig:fig0}). This construction corresponds to the Minkowsky sum of a polyline and a sphere. The mass of the cylindrical elements is lumped at the level of the nodal particles and, at each instant, the fiber shape is defined by the set of positions and orientations of the nodal particles. The density of the fiber $\rho_f$, is assumed to be equal to that of the grains.
The fiber can deform axially and bend, while out-of-plane displacements are blocked to respect the 2D condition assumption. The axial deformation is controlled by a contact normal stiffness $k_a= E_f A_f/l_f$ so that the normal force associated with an elongation (contraction) of a cylindrical element is $F_a=k_a (l-l_f)$, where $l$ is the actual length of the element, $A_f=\pi d^2_f/4$ is the cross section area, $E_f=\eta E_c$ is a material parameter equivalent to an elastic modulus, $\eta$ is a dimensionless coefficient, and $E_c=5\times10^7mg/d^2$ is an elastic parameter used to set the stiffness at the contact scale as described hereafter.     
The bending deformation is controlled by a contact bending stiffness $k_b=E_f I_f/l_f$, where $I_f=\pi d^4_f/64$ is the second moment of area.
A bending moment $M_b=k_b\theta_b$ is associated to a relative rotation $\theta_b$ between two adjacent nodes. %In this framework, the fiber is therefore modeled as an elastic beam.
Details about the implementation of the fiber model are given in \cite{bourrier2013,effeindzourou2016}.
The same approach has been used to study fiber-like object in different contexts such as soil-root interaction \cite{bourrier2013,fakih2019}, mechanical behavior of flexible structures composed of wires or cables \cite{albaba2017,pol2021,effeindzourou2017} and fibers suspension in an fluid \cite{kunhappan2017}.
Another interesting approach to model fibers with DEM is the one presented in \cite{Crassous2023} and adopted to study the behavior of an assembly of twisted frictional fibers \cite{seguin2022twist}.

The particle-particle and particle-cylinder interactions are treated in the same manner assuming a classical spring-dashpot model in the normal direction ($F_n=k_n\delta_n+ \gamma\dot{\delta}_n$ with $k_n=2E_c r_i r_j/\left(r_i+r_j\right)$, where $\delta_n$ is the overlap between the particles, $\gamma=0.3m\sqrt{g/d}$ is a damping coefficient and $r_k$ is the radius of the $k$-$th$ particle) and a spring model in the tangential direction with the elastic tangential displacement between the particles $\delta_t$ bounded with a Coulomb plastic condition with friction coefficient $\mu_c=0.6$ ($F_t=k_t\delta_t\le \mu_cF_n$, $k_t=2k_n/7$). 
The time step is set equal to $2\times10^{-6} \sqrt{d/g}$.
%and $\gamma$ is set in order to have a restitution coefficient of 0.4)

The granular bed is generated by gravity deposition. Once equilibrium is reached, the bed is flattened to the desired height $L_y$ by removing some particles and a relaxation phase under gravity is performed. The fiber is then placed into the granular bed (overlapping particles are removed) with the fiber center at a depth $h$ from the surface of the granular bed. The system is allowed to relax again before starting the simulation; during this phase, the deformation of the fiber is not allowed, in order to start the drag experiment from a perfectly straight fiber configuration. The solid fraction of the granular bed is $\phi\approx0.81$. Finally, we impose a constant displacement rate $v=0.5\sqrt{gd}$ ($Fr \ll1$) to the middle node of the fiber for a total displacement $\Delta x=60d$. It should be noted that the position of the middle node of the fiber along the vertical direction is fixed and equal to $L_y-h$. In the reference configuration, the system dimensions are $L_x=200d$, $L_y=50d$, $h=22d$. A sketch of the system is shown in Figure \ref{fig:fig0}a.

\section{Numerical results}\label{sec:num_results}
\begin{figure}%[h!]
\includegraphics[width =.95 \columnwidth]{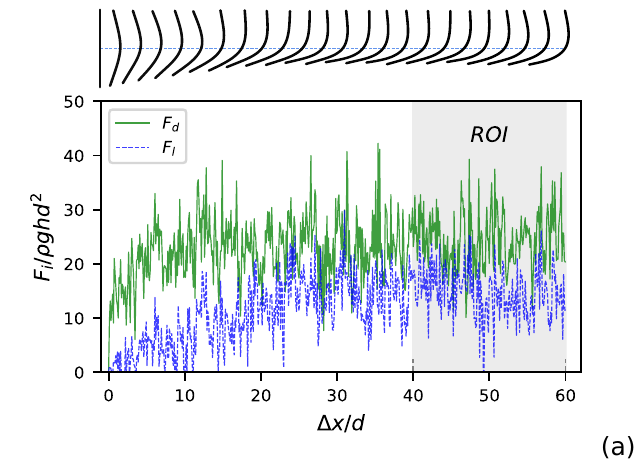}
\includegraphics[width =.95 \columnwidth]{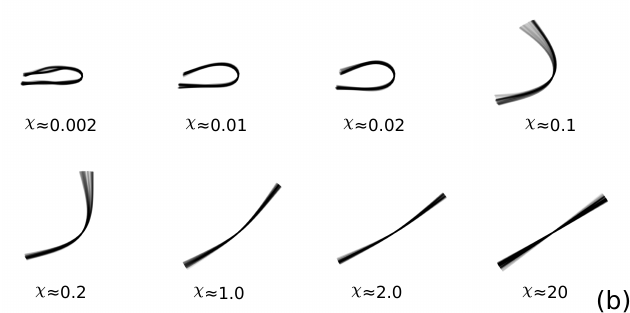}
\caption{(a) Instantaneous drag $F_d$ and lift $F_l$ force on the fiber as a function of the displacement $\Delta x$ (fiber length $L_f=20.5d$ and elastogranular parameter $\chi\approx0.2$ as defined in Eq.~\ref{eq_:chiParam}). The shaded region indicates the region of interest (ROI) where steady state values are computed. At the top of the plot we show the evolution of the shape of the fiber at some steps of the drag test (displacement step of $2.5d$). (b) Typical steady shape of a fiber of length $L_f=20.5d$ for different values of the bending stiffness (the associated elastogranular parameter $\chi$ is reported under each image). Images are a superimposition of instantaneous geometrical configurations of the fiber in the range $40d \leq \Delta x \leq 60d$ (displacement step of $0.05d$).}
\label{fig:fig1}
\end{figure}

In this work, we study the interaction between the fiber and the granular bed by focusing on the total force acting on the fiber, which we decompose into two components: a drag force $F_d$ and a lift force $F_l$. Operationally, we compute the drag (lift) force by summing all the forces acting on the nodal particles of the fiber and projecting them along the drag (orthogonal-to-drag) direction. In Fig.~\ref{fig:fig1} we show the trend of the drag and lift forces during a drag experiment ($L_f=20.5d$, $\eta=0.1$). We also display the evolution of the fiber shape in the same figure.

In the drag experiment, we identify two distinct phases. First, a transient phase, during which the fiber reorients and reconfigures, and the forces acting on it vary in a non-trivial manner. Subsequently, for a sufficiently large displacement $\Delta x$, the fiber adopts a stable geometrical configuration (referred to as the steady shape in what follows), and the system reaches a steady state characterized by drag and lift forces that are, on average, constant. In the context of this work, we focus only on this steady state phase to characterize the interaction between the fiber and the granular bed.
Preliminary analyses have shown that a displacement of $\Delta x = 40d$ is sufficient to reach this steady state, regardless of the fiber length and bending stiffness. All data presented hereafter are therefore obtained by averaging variables over the range $40d \leq \Delta x \leq 60d$ (shaded region in Fig.~\ref{fig:fig1}a). Furthermore, we have verified that the fiber attains the same geometrical configuration, irrespective of its initial orientation.
In this work, we consider fibers of different lengths ($L_f=7.5d$–$30.5d$) and, for each length, use several bending stiffnesses ($\eta$ $= 10^{-3}$–$10^{1}$). In Fig.~\ref{fig:fig1}b, we display some typical steady shapes obtained for a fiber of length $L_f = 20.5d$ and various bending stiffnesses. The elongation of the fiber is always negligible in our study ($<10^{-3}d_f$).
We recall that here we consider a pseudo 2D system, therefore all the quantities are given for a reference out-of-plane thickness $d$. It should be noted that, unlike in the 3D case, particles cannot flow around the fiber, which may enhance their interaction with the fiber.
%---------------------------------------------
%---------------------------------------------

\subsection{Drag force}\label{sec:drag force}
\begin{figure}%[h!]
\includegraphics[width =.9 \columnwidth]{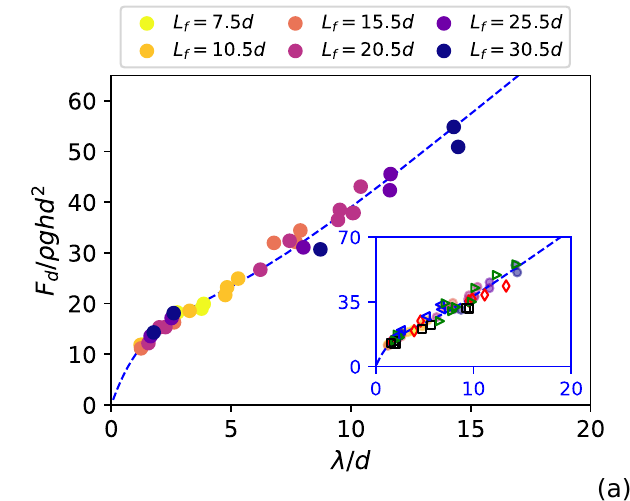}
\includegraphics[width =.9 \columnwidth]{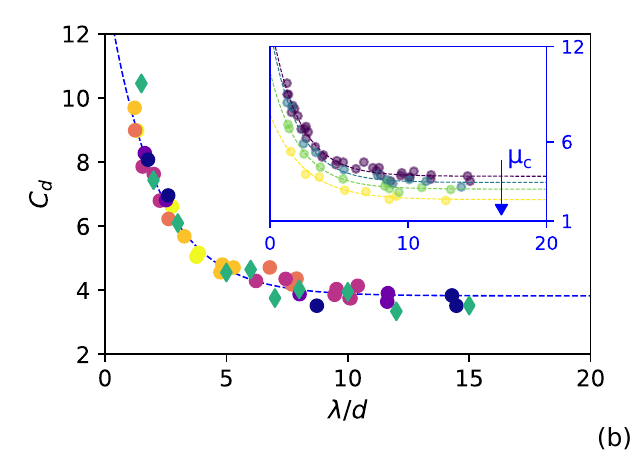}
\caption{(a) Drag force as a function of the characteristic length $\lambda$. Data collapse on the dashed line which corresponds to Eq.~\ref{eq:eq_fd}. Inset: Comparison with data obtained by changing: top pressure $P_{top}=30$-$75 \rho g d$ ($\square$), burial depth $h=47d$ ($\triangleleft$), gravity $2g$ ($\lozenge$), drag velocity $v=0.1$-$1\sqrt{gd}$ ($\triangleright$).
(b) Drag coefficient $C_d$ as a function of the ratio $\lambda/d$ (data are fitted by Eq.~\ref{eq:eq_cd}, $C^{\infty}_d=3.8$, $\alpha_0=2.6$, $\tilde{\lambda}=2$). The diamond markers (\textbf{$\blacklozenge$}) indicates the drag coefficient obtained for a disks of different diameter rescale by the ratio $C^{\infty}_d/C^{\infty}_{d,disk}\approx2.1$.
Inset: Effect of the friction coefficient $\mu_c$ on the drag coefficient ($\mu_c=0.1,~0.2,~0.4,~0.6$).
The legend applies to both figures.
}
\label{fig:fig2} 
\end{figure}

To find a scaling for the drag force, it was natural for us to start from the following hypotheses: (i) the fiber attains a steady shape, (ii) once the fiber has assumed its steady shape, the drag force should scale similarly to the case of a rigid object. It follows that we can express the drag force as  
%---------------------
\begin{equation}
F_d=C_d \rho g h \lambda d%\comm{d~ (\mbox{pseudo-2D)}}
\label{eq:eq_fd}
\end{equation}
%---------------------
where $C_d$ is a (granular) drag coefficient and $\lambda$ a characteristic length of the fiber in the steady shape. This scaling is reminiscent of a frictional criterion, i.e. $F_d$ is proportional to the ambient hydrostatic pressure times a characteristic section, but with a larger coefficient that depends on the frictional properties of the bed (see the inset in Fig.~\ref{fig:fig2}b). Note that $C_d$ can also reasonably vary with the geometry of the system and packing fraction \cite{geng2005,kolb2013,carvalho2022}.  

Using a classical scaling based only on the drag cross section, we observed large deviations from the linear trend predicted by Eq.~\ref{eq:eq_fd}, suggesting that the solely drag cross section is not the relevant length scale in our case.
This started the quest for a characteristic length, and it clearly emerged that a relevant definition for $\lambda$ is
%---------------------
\begin{equation}
 \lambda=\dfrac{S_xS_y}{L_f}
 \label{eq:eq_lambda}
\end{equation}
%---------------------

where $S_x$ and $S_y$
are the projections of the fiber in its steady shape on the drag direction and the perpendicular to the drag direction, respectively (see Fig.~\ref{fig:fig0}b). 
It is interesting to note that, for a disk of diameter $D$, Eq.~\ref{eq:eq_lambda} trivially yields $\lambda = D$, and so our definition is consistent with previous literature results for rigid objects \cite{hilton2013,guillard2014}.
It should be noted that the length $\lambda$ has to be considered only at the steady state when the fiber has found a stable geometrical configuration. During the transient phase, the fiber's shape is constantly evolving towards a steady configuration under the ambient pressure and the actual length $\lambda$ is therefore not meaningful of the fiber-grains interaction, but mainly depends on the initial conditions.
%---------------------------------------------
%---------------------------------------------
In Fig.~\ref{fig:fig2}a, we show that the drag force values collapse onto a single curve when plotted as a function of the characteristic length $\lambda$, independently of the fiber length and bending stiffness. Nevertheless, as $\lambda$ tends to zero, corresponding to very short or extremely flattened fibers, the drag force does not approach zero.  
%To solve this (apparent) contradiction, we extract the drag coefficient from each simulation, i.e. $C_d=F_d/(\rho g h \lambda)$, and report it as a function of the ratio $\tilde{\lambda}=\lambda/d$ in Fig.~\ref{fig:fig2}b. It is clear that when $\lambda$ is comparable to the diameter $d$ of the particles in the granular bed, the drag coefficient increases rapidly as the characteristic size of the fiber decreases. Conversely, for sufficiently large values of $\lambda$, the drag coefficient saturates at a constant value.  
To solve this (apparent) contradiction, we extract the drag coefficient from each simulation, i.e. $C_d=F_d/(\rho g h \lambda d)$, and report it as a function of the characteristic length $\lambda$ in Fig.~\ref{fig:fig2}b. It is clear that when $\lambda$ is comparable to the diameter $d$ of the particles in the granular bed, the drag coefficient increases rapidly as the characteristic size of the fiber decreases. Conversely, for sufficiently large values of $\lambda$, the drag coefficient saturates at a constant value.  
This finite size effect is observed also for a rigid disk as shown in Fig.~\ref{fig:fig2}b (results are obtained replacing the fiber with a disk of diameter $\lambda$). It should be noted that the influence of the size of the intruder on the drag coefficient tends to vanish when it is about 5 times bigger than the particle size for both the disk and the fiber cases. %This effect is probably enhanced by the 2D conditions adopted in this work, but we reasonably expected to observe it also in 3D conditions.
Phenomenologically, this behavior can be attributed to the fact that granular drag originates from the formation and breakage of force chains in the granular skeleton. When the intruder is comparable in size to the particles composing the granular bed, the formation of force chains is governed more by the particle size than by the size of the intruder.
% In this sense, the particle directly in front of the intruder effectively becomes the ``real'' intruder.
%As a consequence, since drag force scales with intruder size in Eq.~\ref{eq:eq_fd}, the proportionality coefficient, i.e. the granular drag coefficient, should be higher for lower $\lambda/d$ ratios.}

The behavior of the drag coefficient $C_d$ with the ratio $\lambda/d$ can be described by an exponential law of the form 

%---------------------
\begin{equation}
% C_d=C^{\infty}_d\left[1+\alpha_0 \exp\left(-\dfrac{\tilde{\lambda}}{\tilde{\lambda}_s}\right)\right]
 C_d=C^{\infty}_d\left[1+\alpha_0 \exp\left(-\dfrac{\lambda}{\tilde{\lambda}d}\right)\right]
 \label{eq:eq_cd}
\end{equation}
%---------------------
where $C^{\infty}_d$ is the value of the drag coefficient when size effects are negligible, $\tilde{\lambda}$ is a characteristic fiber to grain size ratio for which most of the size effects vanish and $\alpha_0$ is a numerical coefficient ($C^{\infty}_d=3.8$, $\alpha_0=2.6$, $\tilde{\lambda}=2$). Including this dependency of the drag coefficient on the $\lambda/d$ ratio into the drag force formulation allows accounting for the discrepancy at small $\lambda$ as shown in  Fig.~\ref{fig:fig2}a.

Before addressing the lift force, we find it useful to briefly comment on the physical implications of the definition of the characteristic length $\lambda$. From Eq.~\ref{eq:eq_lambda}, it follows that the drag force depends linearly not only on the projection of the deformed fiber perpendicular to the drag direction (drag cross section), but also on its projection on the drag direction.

This result is consistent with the experimental observations of Albert et al. \cite{albert2001} which shows that, for a cylindrical body whose axis is aligned with the direction of drag, the drag force increases linearly with the cylinder’s length. Additionally, we have performed a set of simulation using rigid fibers aligned with the drag direction and again observed a linear dependency of the drag force with the fiber length (see Appendix~\ref{sec:appendix}). Another point of agreement with the experimental findings of \cite{albert2001}
is that the dependence on the intruder length along the drag direction surprisingly does not appear to be related to frictional forces along the fiber. In fact, we observed the same dependency even for frictionless fibers (see Appendix~\ref{sec:appendix}). A possible phenomenological explanation is that a longer object may inhibit the collapse of grains behind it, thereby promoting a more stable structure in the granular medium ahead of the object, which results in greater resistance to penetration.

\subsection{Lift force}\label{sec:lift force}
\begin{figure}%[h!]
\includegraphics[width =.9 \columnwidth]{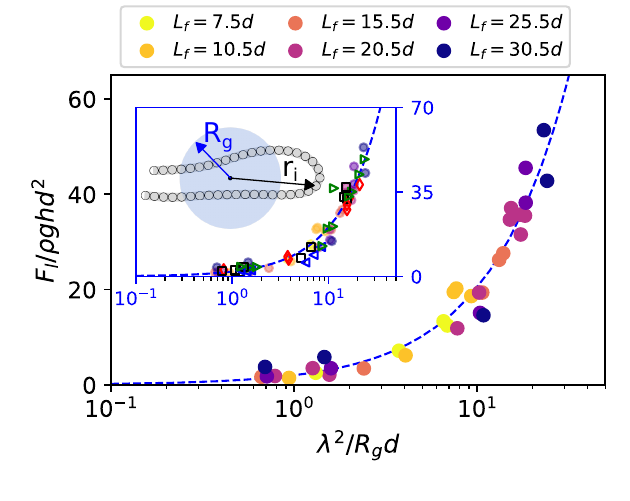}
\caption{Lift force as a function of $\lambda^2/R_g$. Data collapse on the dashed line which corresponds to Eq. \ref{eq:eq_fl} (with $C_l=2$). Inset: Comparison with data obtained changing the main system parameters. Markers are the same as those used in the inset of Fig~\ref{fig:fig2}a.}
\label{fig:fig3}
\end{figure}

In Sec.~\ref{sec:drag force} we have shown that the length $\lambda$ is a relevant length scale for the drag force. Therefore, it seemed natural to us to use the same length when looking for a scaling law for the lift force.
At first, we started from a buoyancy-like scaling inspired by the case of a rigid disk \cite{guillard2014} using $\lambda^2$ as the characteristic surface of the fiber, i.e. $F_l\propto \rho g \lambda^2$. 
We obtained a relative scaling of the data for cases where the fiber is either flattened or remains almost undeformed, but we observed deviations for fibers with intermediate rigidity in which the fiber bends without flattening. This suggests that the sole length scale $\lambda$ is insufficient and that a more refined description of the fiber’s steady shape may be required.
With this in mind, we introduce a second length, the gyration radius $R_g = \sqrt{\frac{1}{N} \sum r^2_i}$, where $r_i$ is the distance of the $i$-th nodal particle from the center of mass of the fiber in its steady shape, and $N$ is the total number of nodal particles (see inset of Fig.~\ref{fig:fig3}). The gyration radius can be interpreted as a measure of the space occupied by the deformed fiber within the granular medium.

Based on this, we propose a scaling law for the lift force of the form
%---------------------
\begin{equation}
F_l=C_l\rho g h \dfrac{\lambda^2}{R_g}d
\label{eq:eq_fl}
\end{equation}
%---------------------

where $C_l$ is the (granular) lift coefficient. From the numerical results, we obtain a constant value of the lift coefficient ($C_l=2$).
In Fig.\ref{fig:fig3}, we show that the scaling proposed in Eq.\ref{eq:eq_fl} enables a good collapse of the lift force data onto a single curve across the entire range of fiber lengths and stiffnesses considered in this study.

At this point, it is worth commenting on the scaling of the lift force. While providing a unique interpretation of Eq.~\ref{eq:eq_fl} is not trivial, we propose two possible readings.
On the one hand, Eq.~\ref{eq:eq_fl} can be interpreted as if the lift force is proportional to the drag force, with the drag-to-lift ratio depending on the shape of the dragged object, here estimated with the ratio $\lambda/R_g$. On the other hand, Eq.~\ref{eq:eq_fl} can be interpreted as if the lift force follows a buoyancy-like criterion, with a lift coefficient depending on the ratio $h/R_g$. This dependency reminds what observed for a rigid disk, for which the lift coefficient was shown to be a function of $h/D$, the burial depth to diameter ratio, before saturating to a constant value for large $h/D$ ratios \cite{guillard2014}.
Finally, it is interesting to note that for all the cases the lift force has a positive value, i.e. the fiber would like to exit the granular bed. 
%---------------------------------------------
%---------------------------------------------
%---------------------------------------------
\section{Discussion}\label{sec:discussion}
%---------------------------------------------
%---------------------------------------------
In Sec.~\ref{sec:num_results}, we have shown that the drag and lift forces can be computed starting from the steady shape of the fiber and, in particular, referring to the characteristic lengths $\lambda$ and $R_g$. It is therefore tempting to try to associate the steady shape of the fiber with the elastic properties of the fiber and the pressure due to the granular environment within which the fiber is moving. 
In this perspective, we define a dimensionless number $\chi$ as the ratio between the bending stiffness of the fiber, considered as an elastic beam, and the moment associated with the ambient (hydrostatic) pressure at the burial depth
%---------------------
\begin{equation}
 \chi=\dfrac{\pi E_f}{32 \rho g h}\dfrac{d_f^3}{L_f^3}
 \label{eq_:chiParam}
\end{equation}
%---------------------
Depending on the value of this elastogranular parameter, it is possible to distinguish three macro regimes for the deformation of the fiber: 
\begin{itemize}
    \item $\chi<10^{-2}$, the fiber largely deforms and eventually completely flattens under the granular pressure;
    \item $10^{-2}<\chi<1$, the fiber bends under the granular pressure and reconfigures interacting with the granular medium;
    \item $\chi > 1$, the fiber negligibly deforms under the granular pressure and  behaves similarly to a rigid rod.
\end{itemize}

\begin{figure}%[h!]
\includegraphics[width =.9 \columnwidth]{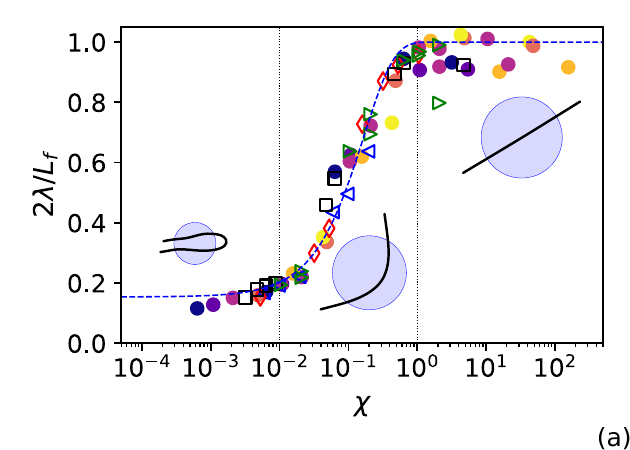}
\includegraphics[width =.9 \columnwidth]{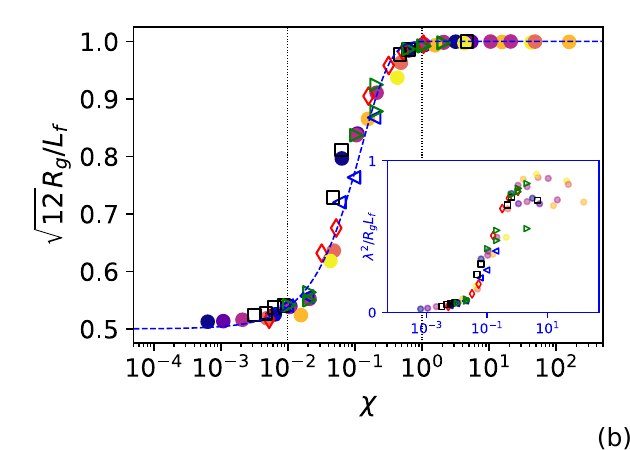}
\caption{(a) Characteristic length $\lambda$ as a function of the elastogranular parameter $\chi$. Dashed line corresponds to Eq.~\ref{eq:eq_lamChi} ($\alpha_{\lambda}=0.85$, $\chi_{_{\lambda}}=0.17$). Three steady-shapes typical of the elastogranular regimes are displayed inside the plot. The circular filled area around the fibers has radius $R_g$ and is centered on the center of mass of the deformed fiber. (b) Gyration radius $R_g$ as a function of $\chi$. Dashed line corresponds to by Eq. \ref{eq:eq_RcChi} ($\chi_{_{R_g}}=0.13$) Inset: Ratio $\lambda^2/R_g$ as a function of the rigidity parameter $\chi$.
Markers are the same as those used in the inset of Fig~\ref{fig:fig2}a for all the figures.}
\label{fig:fig4}
\end{figure}

In this sense, the parameter $\chi$ can be interpreted as a rigidity parameter that is reasonably associated with the steady shape of the fiber and, to some extent, governs the interaction regime between the fiber and the granular medium.
To support this statement, we show in Fig.~\ref{fig:fig4}a the characteristic length $\lambda$ as a function of the parameter $\chi$. The data collapse on a single master curve, which proves that $\chi$ is a relevant parameter to describe the steady shape of the fiber in our system. The three regimes mentioned above are visible in Fig.~\ref{fig:fig4}a where three typical steady shapes for the fiber are displayed (see also Fig.~\ref{fig:fig1}b). To describe the variation of $\lambda$ with $\chi$ we propose the following model
%---------------------
\begin{equation}
 %\dfrac{2\lambda}{L_f}=1-\alpha_{\lambda}\exp\left(-\dfrac{\chi}{\chi_{_{\lambda}}}\right)
 \lambda=\left[1-\alpha_{\lambda}\exp\left(-\dfrac{\chi}{\chi_{_{\lambda}}}\right)\right]\dfrac{L_f}{2}
 \label{eq:eq_lamChi}
\end{equation}
%---------------------
where $\alpha_{\lambda}$ is a numerical coefficient and $\chi_{_{\lambda}}$ is a characteristic value of the elastogranular parameter for which effects on $\lambda$ tend to vanish ($\alpha_{\lambda}=0.85$, $\chi_{_{\lambda}}=0.17$).%, i.e. the fiber tends to behave as a rigid rod.
%---------------------------------------------
%---------------------------------------------
Similarly, we observe that the gyration radius $R_g$ also scales with $\chi$ as reported in Fig.~\ref{fig:fig4}b. To describe the variation of $R_g$ with $\chi$ we propose the following model 
%---------------------
\begin{equation}
 R_g=\left[1-0.5\exp\left(-\dfrac{\chi}{\chi_{_{R_g}}}\right)\right]\dfrac{L_f}{\sqrt{12}}
 \label{eq:eq_RcChi}
\end{equation}

%---------------------
where the numerical coefficients can be obtained by considering that for a perfectly straight fiber (case $\chi \rightarrow \infty$) $R_g=L_f/\sqrt{12}$ and for a fiber perfectly folded in half (case $\chi \rightarrow 0$) $R_g$ tends to $L_f/2\sqrt{12}$. Finally,  $\chi_{_{R_g}}$ is a characteristic value of the elastogranular parameter for which effects on $R_g$ tend to vanish ($\chi_{_{R_g}}=0.13$). We would like to emphasize that the lift and drag forces at steady state can be determined simply from the knowledge of the parameter $\chi$. 

The predictions from Eq.~\ref{eq:eq_fd} and Eq.~\ref{eq:eq_fl}, assuming a variation of the parameter $\chi$ within the range considered in this study, are shown with a dashed blue line in Fig.~\ref{fig:fig2}a and Fig.~\ref{fig:fig3} respectively. We observe an excellent agreement between analytical predictions and simulation data.
The force scalings are finally tested against numerical data obtained by changing the main system's parameters: gravity ($2g$), drag velocity ($v=0.1$-$1\sqrt{gd}$), burial depth ($h=47d$). Note that the burial depth was changed by using a granular bed of greater thickness ($L_y=75d$, see Fig.~\ref{fig:fig0}c).
Additionally, the ambient pressure at the fiber level was modified by imposing a top pressure $P_{top}=30$-$75 \rho g d$ through a rigid heavy wall on the top of the granular bed (see Fig.~\ref{fig:fig0}c), which mimics the effect of a greater burial depth $h=P_{top}/\left(\rho g\right)$ . All the data collapse on a single master curve proving the robustness of the proposed scalings (see insets in Fig.~\ref{fig:fig2}a and Fig.~\ref{fig:fig3}).

%---------------------------------------------
\begin{figure}%[h!]
\includegraphics[width =.9 \columnwidth]{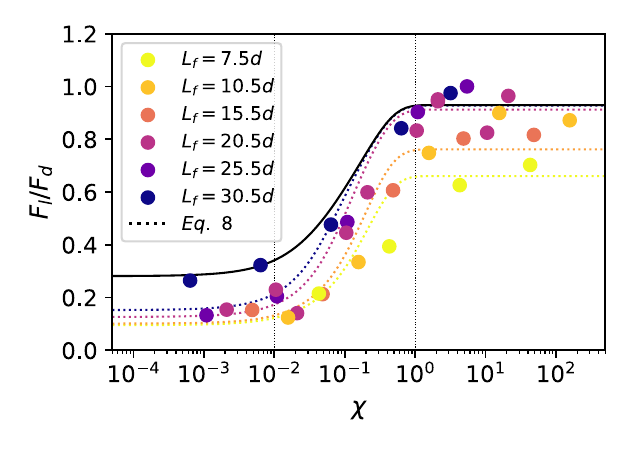}
\caption{Lift-to-drag ratio as a function of the rigidity parameter $\chi$. Predictions of Eq. \ref{eq:lift2drag} are shown with dashed lines (line color identify the associated fiber length $L_f$). The solid black line represents the very long fiber limit when the lift-to-drag ratio is no more dependent on the fiber length.}
\label{fig:fig6}
\end{figure}
%---------------------------------------------

Finally, it is interesting to note that the proposed models for the drag and lift force trivially give an expression for the drag-to-lift ratio 
\begin{equation}\label{eq:lift2drag}
    \dfrac{F_l}{F_d}=\dfrac{C_l}{C_d}\dfrac{\lambda}{R_g}
\end{equation}
Note that, again, the lift-to-drag ratio can be computed starting only from the elastogranular parameter $\chi$.
In Fig.~\ref{fig:fig6} we report the values of the lift-to-drag ratio as a function of $\chi$. We also report the predictions of Eq.~\ref{eq:lift2drag} that show a qualitative agreement with the numerical data.
Globally, the lift-to-drag ratio increases with the parameter $\chi$. It starts from a nonzero minimum for highly deformable fibers ($\chi < 10^{-2}$), increases monotonically for intermediate values of the elastogranular parameter ($10^{-2}<\chi <1$), and saturates to a constant value in the limit of pseudo-rigid fibers ($\chi > 1$).
Interestingly, we note that the limiting values of the lift-to-drag ratio appear to depend on fiber length for short fibers: longer fibers exhibit higher values. 
This behavior may arise from the dependence of the drag coefficient on the ratio $\lambda/d$ (see Eq.~\ref{eq:eq_cd}), indicating size effects associated with the granular nature of the environment within which the fiber moves. For very long fibers, even when they tend to flatten, the characteristic length $\lambda$ remains larger than the particle size of the granular bed. Similarly, in the limit of pseudo-rigid fibers, it is straightforward to observe that the characteristic length $\lambda$ is larger for longer fibers. In both cases, this leads to a higher drag coefficient for shorter fibers. Since we assumed here the lift coefficient to be independent of fiber size, longer fibers exhibit a higher lift-to-drag ratio.
It is worth noting that, for fibers significantly larger than the grain size, the lift-to-drag ratio becomes independent of fiber length ($0.3 \lesssim F_l/F_d \lesssim 0.95$, solid line in Fig.~\ref{fig:fig6}), as size effects disappear, as expected.

\begin{figure}%[h!]
\includegraphics[width =.9 \columnwidth]{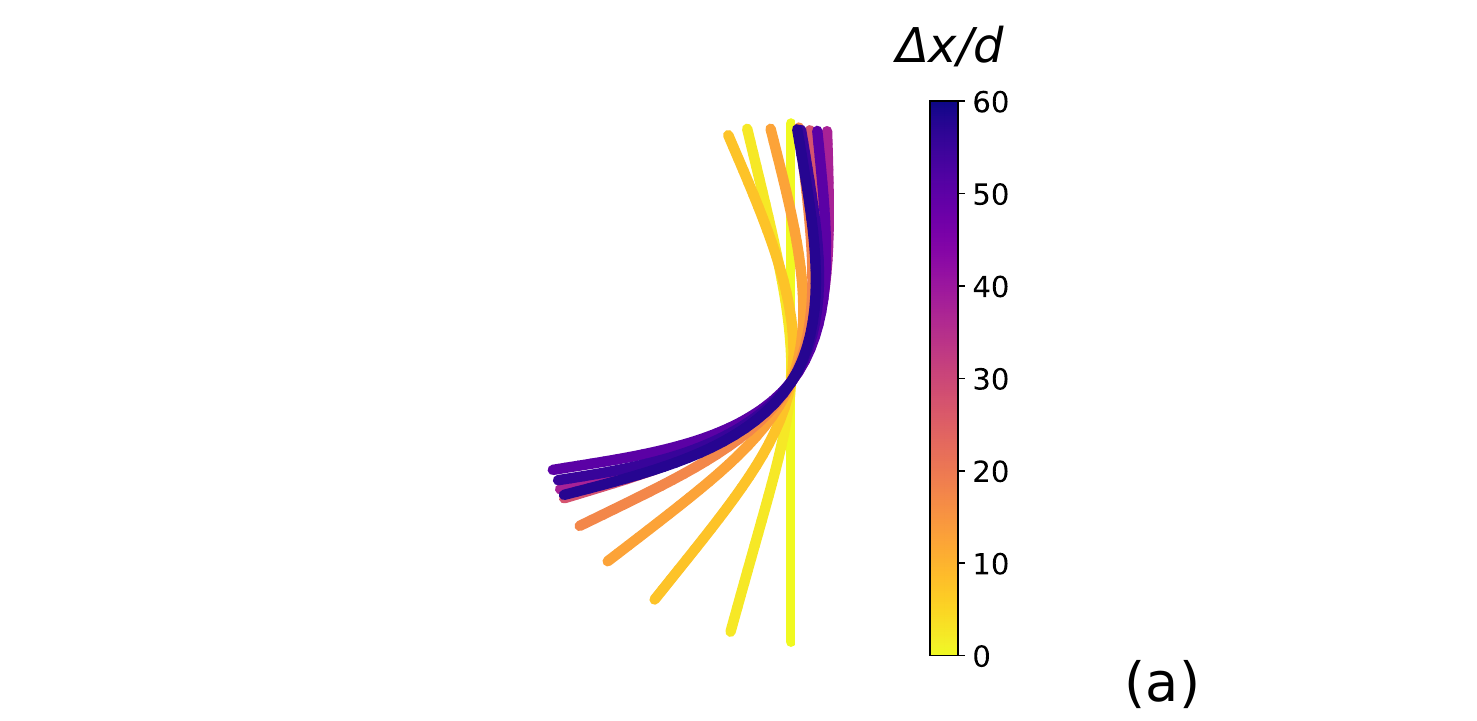}
\includegraphics[width =.9 \columnwidth]{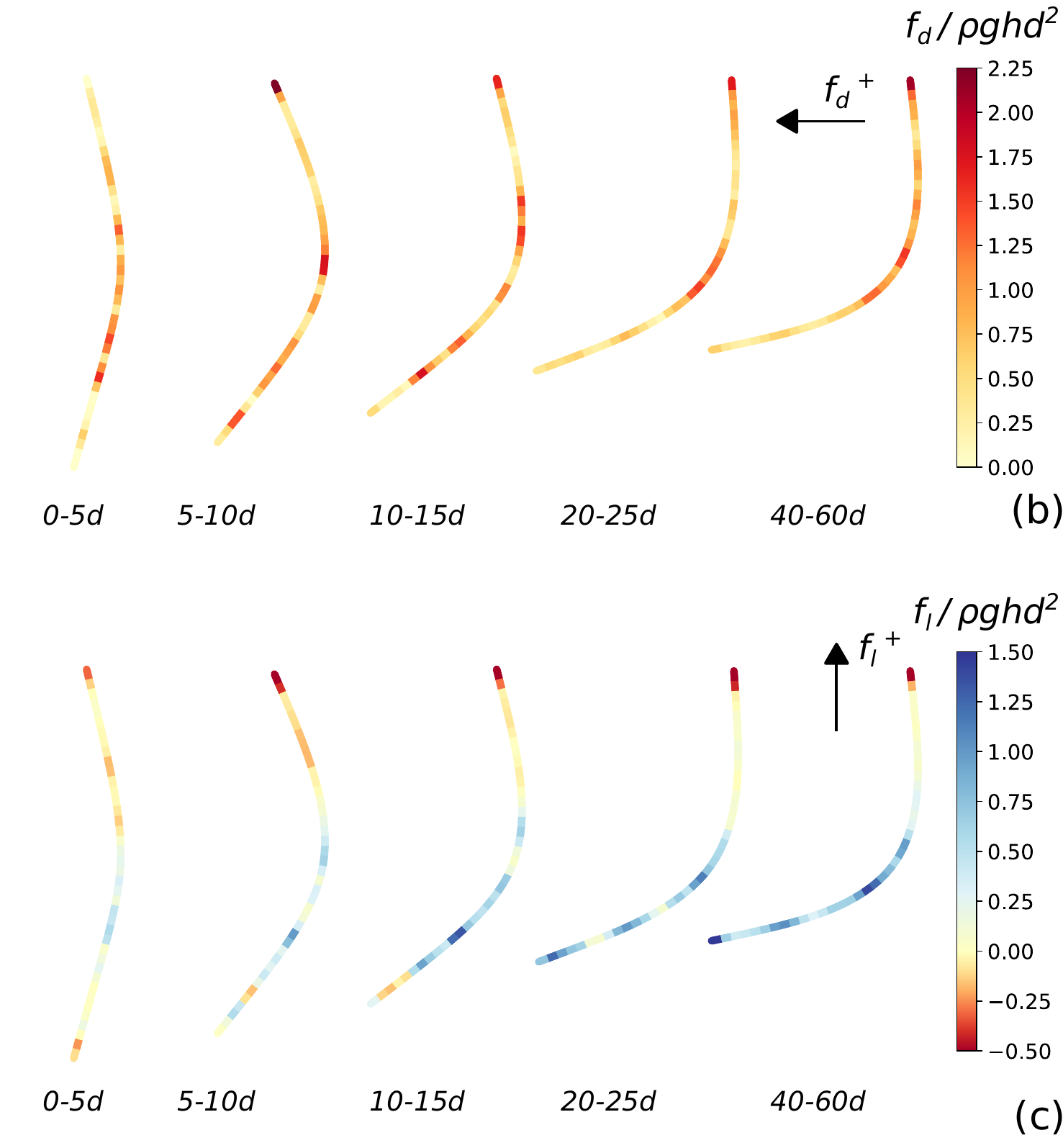}
\caption{(a) Evolution of the fiber shape with the fiber displacement $\Delta x$ (case $L_f=20.5d$ and $\chi\approx0.2$). Distribution of: (b) the local drag $f_d$ and (c) the local lift $f_l$ along the fiber at different intervals of the displacement $\Delta x$. The forces are obtained from a time average of the local force on each cylindrical element composing the fiber in the considered displacement range. The sign convention for the local forces is precised in the figure.}
\label{fig:fig7}
\end{figure}

Finally, we would like to discuss the shape of the fiber in the steady state.
It is possible to distinguish three typical shapes which are characteristic of the three regime above mentioned: a half flattened fiber, a bent fiber and an undeformed one. An example of these shapes is given by the cases with $\chi\approx0.002,~\chi\approx0.2$ and $\chi\approx20$ in Fig.~\ref{fig:fig1}b, respectively.
The first case occurs when the fiber is highly deformable relative to the granular environment ($\chi < 10^{-2}$). In this regime, when dragged across the medium, the fiber simply flattens by folding in half under the granular pressure.
The second case, in which the fiber undergoes finite bending ($10^{-2} < \chi < 1$), is more interesting, as the steady shape results from the interplay between the fiber’s elastic properties and the granular pressure acting on it. An example of the fiber’s shape evolution is shown in Fig.~\ref{fig:fig7}a (case $L_f=20.5d$, $\chi\approx0.2$). In this regime, the fiber may exhibit markedly different shapes above and below the pulling point, which we argue can be rationalized in terms of the local angle formed by the top and bottom portions of the fiber with respect to the drag direction (referred to as the angle of attack in what follows) and its effect on the local pressure. To support our statement we show in Fig.~\ref{fig:fig7} the evolution of the fiber shape with the displacement $\Delta x$ and the distributions of the local drag $f_d$ and lift $f_l$ forces along the fiber. The local forces are computed by performing a time average of the instantaneous local force acting on each single cylindrical element composing the fiber in the associated displacement range.
At first, the fiber bends under the horizontal pressure and develops different angles of attack above and below the pulling point. In this phase, the fiber will reasonably bend more below the pulling point due to the higher horizontal pressure ($\Delta x \leq 10d$ in Fig.~\ref{fig:fig7}b). Subsequently, the fiber begins to experience a local vertical pressure that differs in both magnitude and direction above and below the pulling point, as it strongly depends on the local angle of attack (see Appendix~\ref{sec:appendix}). Indeed, we observe that the vertical pressure is much higher in the bottom part of the fiber than in the top one (see Fig.~\ref{fig:fig7}c for $\Delta x > 10d$). 
This vertical pressure difference induces a clockwise rotation of the fiber, which seeks an equilibrium position by adjusting its angle of attack in the top part (see Fig.~\ref{fig:fig7}). 
The evolution of the fiber shape illustrates how the fiber adapts within the granular environment to reach a steady configuration by adjusting its shape and local angle of attack, thereby highlighting the strong elastogranular interaction.

%From Fig.~\ref{fig:fig7}b it is evident that the local lift force is much greater below the pulling point, causing a higher bending in the bottom part of the fiber than in the top one (note that also the direction of the force is different).
%The local drag force is instead a bit higher above the pulling point than below, which results again from an angle of attack effect.
%
%
%
Finally, in the case of a nearly rigid fiber, the fiber simply rotates until it reaches a preferred orientation in the range $35^\circ$–$45^\circ$. Interestingly, this range corresponds to the attack angles at which the lift force is maximized (see Appendix~\ref{sec:appendix}).
%However, we do not have an explanation for the origin of this particular orientation.}
%\comm{metterei piuttosto che non è oggetto di questo sudio, oppure nelle conclusioni direi che potrà essere interessante investigare l'origine di questo effetto che è simile a quello sperimentato nei fluidi}

%---------------------------------------------
%---------------------------------------------

%---------------------------------------------
%---------------------------------------------
\section{Conclusion}\label{sec:conclusion}
%---------------------------------------------
%---------------------------------------------
In this work, we investigated numerically the behavior of drag and lift forces on a flexible fiber moving within a granular medium.
We found that, after a large enough displacement, the system reaches a steady state in which the fiber shape does not vary and the forces acting on it are, on average, constant. Under these conditions, we defined two characteristic lengths, $\lambda$ and $R_g$, and showed how they allow definition of unique scaling laws for the drag and lift forces which apply from extremely deformable to pseudo-rigid fibers.  

We proposed a dimensionless parameter, the elastogranular parameter $\chi$, which controls the fiber-grains interaction %defined as the ratio between fiber bending stiffness and the moment associated to the ambient (hydrostatic) pressure, 
and showed how the characteristic lengths, and so the drag and lift forces, can be expressed as an only function of this dimensionless parameter.
This is particularly interesting because of the elastogranular parameter can be computed only from the geometry of the system and the mechanical properties of the fiber, which are known a priori.

Our findings provide a first step toward a better understanding of the interaction between fibers and granular media by proposing simple scaling laws for the forces acting on a fiber, and by showing that the problem is governed by a single dimensionless parameter. Knowledge of the forces at the fiber scale is, in fact, a fundamental ingredient for rheological models attempting to describe the behavior of fiber–grain systems. %\cite{darbois2023,wierzchalek2025}.
Despite the progress made, several aspects still require further investigation, such as the potential saturation of the lift force at large burial depths and whether similar dimensional arguments, with the addition of a kinetic pressure term, apply also in the high Froude number limit.
We also plan to extend the model to 3D conditions, where the interaction between the fiber and the particles may differ because particles can flow around the fiber, and to explore additional configurations such as a fiber interacting with a granular flow. 
In the future, it will be interesting to compare the numerical approach here adopted with experimental results of elastogranular interaction \cite{seguin2018,algarra2018,schunter2018}.
Finally, an interesting direction for future work is to compare the forces measured in our study with predictions from resistive force theory (RFT) \cite{li2013terradynamics}, to test the RFT when the size of the intruder is comparable to the grain size.

%----------------------------------------------------------

%-----------------------------
%\begin{table}[b]
%\caption{\label{tab:table4}%
%Characteristic of the fibers considered in the simulations}
%\begin{ruledtabular}
%\begin{tabular}{ccddd}
%One&Two&
%\multicolumn{1}{c}{\textrm{Three}}&
%\multicolumn{1}{c}{\textrm{Four}}&
%\multicolumn{1}{c}{\textrm{Five}}\\
%\mbox{Three}&\mbox{Four}&\mbox{Five}\\
%\hline
%one&two&\mbox{three}&\mbox{four}&\mbox{five}\\
%He&2& 2.77234 & 45672. & 0.69 \\
%C\footnote{Some tables require footnotes.}
%  &C\footnote{Some tables need more than one footnote.}
%  & 12537.64 & 37.66345 & 86.37 \\
%\end{tabular}
%\end{ruledtabular}
%\end{table}

%-----------------------------------------------------------

%---------------------------------------------
%-----------------------------------------------------------
%-----------------------------------------------------------
\begin{acknowledgments}
A.P. acknowledges the support of French National Research Agency (Grant No. ANR-23-CE30-0032 ConFiG) and I-SITE Excellence Program (University of Montpellier). S.S. acknowledges the financial support of the Erasmus+ Programme of the European Union through the  Erasmus+ mobility for traineeships project for higher education students (Grant Agreement No. 2024-1-IT02-KA131-HED-000199330).
A.P. is thankful to Patrick Richard for fruitful discussions and careful reading of the manuscript.
\end{acknowledgments}

%%%%%%%%%%%%%%%%%%%%%%%%%%%%%%%%%%%%%%%%%%%%%%%%%%%
\appendix
%%%%%%%%%%%%%%%%%%%%%%%%%%%%%%%%%%%%%%%%%%%%%%%%%%%
\counterwithin{figure}{section}
\counterwithin{table}{section}
\renewcommand\thefigure{\thesection\arabic{figure}}
\renewcommand\thetable{\thesection\arabic{table}}
%%%%%%%%%%%%%%%%%%%%%%%%%%%%%%%%%%%%%%%%%%%%%%%%%%%
%\section{Effect of the angle of attack on drag and lift forces}
\section{Effect of the orientation and length of the intruder}
\label{sec:appendix}
\begin{figure}[h!]
\includegraphics[width =.95 \columnwidth]{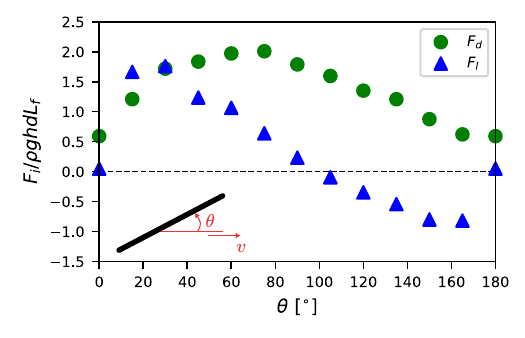}
\caption{ Drag $F_d$ and lift $F_l$ force for a perfectly rigid fiber as a function of the angle of attack $\theta$ with respect ot the drag direction ($L_f=20.5d)$.}
\label{fig:figApp}
\end{figure}

%%%%%%%%%%%%%%%%%%%%%%%%%%%%%%%%%%%%%%%%%%%%%%%%%%%
To gain insight into the effect of the angle of attack on the drag and lift forces, we performed a set of simulations using the same setup described in Sec.~\ref{sec:methodology}, but this time with a perfectly rigid fiber of length $L_f = 20.5d$ and with a fixed orientation $\theta$ as the intruder.
The results are shown in Fig.~\ref{fig:figApp}, where the force values are computed as time averages after a sufficient initial displacement to ensure a steady state is reached.
We note that, even when the intruder size is comparable to the particles composing the granular medium, the forces acting on the intruder are strongly influenced by the angle of attack. The same trend observed for macroscopic intruders is recovered \cite{ding2011drag, zhang2014}, with a marked asymmetry about $\theta = 90^{\circ}$: both drag and lift forces are larger in magnitude for $0^{\circ} \leq \theta \leq 90^{\circ}$ than for $90^{\circ} \leq \theta \leq 180^{\circ}$. The drag force reaches its maximum when the angle of attack is around $70^{\circ}$, while the lift force peaks at approximately $30^{\circ}$.
%---------------------------------------------
%---------------------------------------------

\begin{figure}%[h!]
\includegraphics[width =.95 \columnwidth]{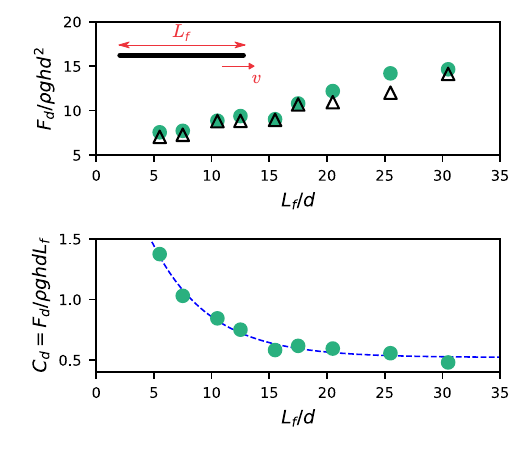}
\caption{(a) Drag force $F_d$ as a function of the fiber length $L_f$ for a fiber oriented along the drag direction. The case of a frictionless fiber is also reported ($\triangle$).
(b) Drag force rescaled by the fiber's length as a function of the ratio $L_f/d$. Data are fitted by Eq.~\ref{eq:eq_cd} ($C^{\infty}_d=0.5$, $\alpha_0=4.9$, $\tilde{\lambda}=4.9$).}
\label{fig:figApp2}
\end{figure}

From the definition of the characteristic length $\lambda$ given in Sec.~\ref{sec:drag force}, it follows that the drag force depends linearly on the length of the intruder along the drag direction. To support this result, we performed a set of simulations considering rigid fibers of different lengths $L_f$ but with the same diameter $d_f$, oriented at $\theta=0^{\circ}$.
In Fig.~\ref{fig:figApp2}a we clearly observe that the drag force increases linearly with the fiber's length, confirming the role of the length of the intruder along the drag direction. A slight deviation from the linear trend can be seen for the smaller values of $L_f$, which can be associated with size effects that emerge when the fiber length is comparable to the particle diameter. This behavior is similar to what was observed for the case of a rigid disk and for a deformable fiber in Sec.~\ref{sec:drag force}. Indeed, when computing an equivalent of the drag coefficient by rescaling the drag force with a reference force associated with the fiber length ($F_d/\rho g h d L_f$), the data are well fitted by Eq.~\ref{eq:eq_cd} (see Fig.~\ref{fig:figApp2}). It is interesting to note that, even in the case of a frictionless contact between the fiber and the particles (see Fig.~\ref{fig:figApp2}), a linear trend is observed, in agreement with experimental findings \cite{albert2001}. This suggests that the dependence of the drag force on the length of the intruder along the drag direction has a geometric origin.  

%---------------------------------------------
%\nocite{*}

\newpage

\bibliography{apssamp}% Produces the bibliography via BibTeX.

\end{document}